\documentclass[useAMS]{mn2e}
\usepackage{url}
\usepackage{graphicx}
\usepackage{soul}
\usepackage[T1]{fontenc}

\title[]{Enhanced activity of the Southern Taurids in 2005 and 2015} 
\author[A. Olech et al.]{A. Olech$^{1}$\thanks{e-mail:
olech@camk.edu.pl},  
P. \.Zo{\l}\k{a}dek$^2$, M. Wi\'sniewski$^{2,3}$, Z. Tymi\'nski$^4$, M. Stolarz$^2$, M. B\k{e}ben$^2$, 
\newauthor D. Dorosz$^2$, T. Fajfer$^2$, K. Fietkiewicz$^2$, M. Gawro\'nski$^5$, M. Gozdalski$^2$,
\newauthor M. Ka{\l}u\.zny$^2$, M. Krasnowski$^2$, H. Krygiel$^2$, T. Krzy\.zanowski$^2$, M. Kwinta$^2$,
\newauthor T. {\L}ojek$^2$, M. Maciejewski$^2$, S. Miernicki$^2$, M. Myszkiewicz$^2$, P. Nowak$^2$, 
\newauthor K. Polak$^2$, K. Polakowski$^2$, J. Laskowski$^2$, M. Szlagor$^2$, G. Tissler$^2$, 
\newauthor T. Suchodolski$^6$, W. W\k{e}grzyk$^2$, P. Wo\'zniak$^2$ and P. Zar\k{e}ba$^2$\\
$^{1}$Nicolaus Copernicus Astronomical Center,
Polish Academy of Sciences, ul.~Bartycka~18, 00-716~Warszawa, Poland\\
$^{2}$Comets and Meteors Workshop, ul. Bartycka 18, 00-716 Warszawa, Poland\\
$^{3}$ Central Office of Measures, ul. Elektoralna 2, 00-139 Warsaw, Poland\\
$^{4}$ Narodowe Centrum Bada\'n J\k{a}drowych, O\'srodek Radioizotop\'ow POLATOM,
ul. So{\l}tana 7, 05-400 Otwock, Poland\\
$^{5}$ Toru\'n Centre for Astronomy, Faculty of Physics, Astronomy and Applied Informatics, 
N. Copernicus University,\\  ul. Grudzi\c{a}dzka 5, 87-100 Toru\'n, Poland\\
$^{6}$ Space Research Centre, Polish Academy of Sciences, ul. Bartycka 18A, 00-716 Warszawa, Poland
}

\begin{document}

\date{Accepted ......, Received ...........; in original form 2016 December 22}

\pagerange{\pageref{firstpage}--\pageref{lastpage}} \pubyear{2017}

\maketitle

\label{firstpage}

\begin{abstract}

The paper presents an analysis of Polish Fireball Network (PFN)
observations of enhanced activity of the Southern Taurid meteor shower
in 2005 and 2015. In 2005, between October 20 and November 10, seven
stations of PFN determined 107 accurate orbits with 37 of them belonging
to the Southern Taurid shower. In the same period of 2015, 25 stations
of PFN recorded 719 accurate orbits with 215 orbits of the Southern
Taurids. Both maxima were rich in fireballs which accounted to 17\% of
all observed Taurids. The whole sample of Taurid fireballs is quite
uniform in the sense of starting and terminal heights of the trajectory.
On the other hand a clear decreasing trend in geocentric velocity 
with increasing solar longitude was observed.

Orbital parameters of observed Southern Taurids were compared to orbital
elements of Near Earth Objects (NEO) from the NEODYS-2 database. Using
the Drummond criterion $D'$ with threshold as low as 0.06, we found over
100 fireballs strikingly similar to the orbit of asteroid 2015 TX24.
Several dozens of Southern Taurids have orbits similar to three other
asteroids, namely: 2005 TF50, 2005 UR and 2010 TU149. All mentioned NEOs
have orbital periods very close to the 7:2 resonance with Jupiter's
orbit. It confirms a theory of a "resonant meteoroid swarm" within the
Taurid complex that predicts that in specific years, the Earth is hit by
a greater number of meteoroids capable of producing fireballs. 

\end{abstract}

\begin{keywords}
meteorites, meteors, meteoroids, asteroids
\end{keywords}

\begin{table*}
\centering
\caption[]{Basic data on the PFN stations which observed Taurid shower maximum in 2005.}   
\begin{tabular}{|l|l|c|c|c|l|l|}
\hline
\hline
Code & Site & Longitude [$^\circ$] & Latitude [$^\circ$] & Elev. [m] & Camera & Lens \\
\hline
PFN03 & Z{\l}otok{\l}os & 20.9086 E & 52.0062 N & 128 & Tayama C3102-01A1 & Ernitec 4mm f/1.2 \\
PFN05 & Pozna\'n & 16.9098 E & 52.4280 N & 89 & Tayama C3102-01A1 & Ernitec 4mm f/1.2 \\
PFN06 & Krak\'ow & 19.9424 E & 50.0216 N & 250 & Mintron MTV-23X11C & Ernitec 4mm f/1.2 \\
PFN09 & \.Zabik\'ow & 22.5498 E & 51.8068 N & 154 & Praktica L2 & Vivitar 28mm f/2.5 \\
PFN13 & Toru\'n & 18.6209 E & 53.0252 N & 66 & Siemens CCBB1320 & Ernitec 4mm f/1.2 \\
PFN17 & Gdynia & 18.5473 E & 54.5069 N & 34 & Mintron MTV-23X11C & Ernitec 4mm f/1.2 \\
PFN19 & Kobiernice & 19.2018 E & 49.8377 N & 345 & Mintron MTV-23X11C & Ernitec 4mm f/1.2 \\
\hline
\hline
\end{tabular}
\end{table*}

\section{Introduction}

The Taurids are two ecliptic meteor showers active every autumn. They
have been observed since the second half of 19th century and they were
examined using photographic methods for the first time in the  mid-20th
century (Whipple 1940, Wright and Whipple 1950). At that time they were
divided into two segments, the Northern Taurids and the Southern
Taurids, and now they are recognized as two separate showers. During
those studies the Taurids were found out to be a part of a bigger
complex which includes also smaller swarms with ecliptic radiants
positioned nearby. A very complicated structure of the Taurids complex
along with a long period of their activity make the problem of their
origin and evolution anything but trivial. Actually it is thought that
the 2P/Encke comet is the source and parent body of the Taurids (Whipple
1940), however both 2P/Encke and Taurids might be the remnants of a much
larger object, which has disintegrated over the past 20000 to 30000
years (Clube and Napier 1984, Asher et al. 1993, Babadzhanov et al. 2008).

A wide analysis of the Taurids swarm based on studies of photographic
orbits available in the IAU database was conducted Porub\v{c}an et al.
(2006). While examining orbital similarities among the Taurids
registered in photographs the authors singled out 15 filaments belonging
to the Taurid complex. For 7 substreams they found 9 Near Earth Objects
(NEOs) which can be connected to them. Among those worth listing there
are the 2004 TG10 (similarity to the Northern Taurids) and the
2005 TF50 situated near 7:2 resonance with Jupiter. The
conducted retrograde orbit integration shows a possibility of common
source of many discovered filaments which date of creation is thought to
be about 4500 years ago.

Most recently, analysis of Porub\v{c}an et al. (2006) was followed by
Jopek (2011) who identified as many as 14 possible parent bodies of the
Taurids stream. Additionally, Ka\v{n}uchov\'a and Svore\v{n} (2014) used
the method of indices to study the autumn part of the Taurid complex and
identified as many as 13 associations.

The analysis of the activity of the Taurids over decades and centuries
is a matter of separate studies. Usually that swarm is characterized by
Zenithal Hourly Rate not exceeding 5 but while studying their activity
for last several decades the scientists managed to notice years in which
the activity of the Taurids was higher than average; particularly they
managed to connect the noticeably higher activity of fireballs with the
swarm of the Taurids (Asher 1991, Asher and Clube 1993, Asher and Izumi
1998). The increased activity of the shower was connected with the
existence of a stream of particles which remains in 7:2 resonance with
Jupiter. An increased activity of the swarm was predicted for years
1995, 1998, 2005 and 2008. 

The return in 2005 was spectacular with both enhanced global activity
and maximum rich in fireballs (Dubietis and Arlt, 2006). In 2008 the
activity of the shower was lower but still it may be considered as
enhanced (Jenniskens et al. 2008, Shrben\'y and Spurn\'y 2012).

According to the Asher's model, the next swarm encounter year was
expected in 2015. Other points of meeting with the 7:2 resonance stream
were computed and the increased activity was predicted also for years
2019, 2022, 2025, 2032 and 2039.

In the following paper we described the peaks of the Taurids' activity
observed in 2005 and 2015. According to predictions, in 2015 the observed
maximum was characterized by a high number of bright fireballs. Among
them there were two very bright events which detailed description might
be found in a separate paper (Olech et al. 2016). Many orbits of
fireballs were found similar to the orbits of the 2005 UR and the 2005
TF50 objects (Olech et al. 2016) as well to the asteroid 2015 TX24 
(\.Zo{\l}\k{a}dek et al. 2016). In the following paper we are examining
the activity of the Taurids and their orbital similarities to NEO
objects for years 2005 and 2015.

\section{Observations}

The PFN is the project whose main goal is regularly monitoring the sky
over Poland in order to detect bright fireballs occurring over the whole
territory of the country (Olech et al. 2006, Wi\'sniewski et al. 2016).
It is kept by amateur astronomers associated in Comets and Meteors
Workshop (CMW) and coordinated by astronomers from Copernicus
Astronomical Center in Warsaw, Poland. Currently, there are 35 fireball
stations belonging to PFN that operate during each clear night. It total
over 70 sensitive CCTV cameras with fast and wide angle lenses are used.

\subsection{Observations in 2005}

The PFN was created a bit over one year earlier so at the end of 2005 it
consisted of less than 10 active cameras positioned in different parts
of the country. The Tayama C3102-01A1 cameras combined with the Ernitec
1.2/4mm lens were the basic set for video observations - one of the
faster non-enhanced video sets in those times, allowing to observe
meteors up to $+2$ magnitude. Also other cameras with similar parameters
- the Mintron MTV-23X11C and the Siemens CCBB 1320 - were used. During
the analysis of the Taurids in 2005 and 2015 we focused on solar
longitudes ($\lambda_\odot$) from 206 to 228 degrees (from
October 20 to November 10). In 2005, for the aforementioned period, the
weather conditions were changeable. From October 23 to 25 we had no data
due to bad weather conditions and then it was repeated on November 5 and
7-8. Between 2005 October 26 and November 4 the weather conditions at
night were quite good for a change. The New Moon on November 2 offered
good conditions for observations of not only fireballs but also fainter
meteors.

In the Table 1 we present a list of stations observing the peak activity
of the Taurids in 2005. These observations were conducted using video
sets described above and the output was controlled by the MetRec
software (Molau 1999). Additionally, we conducted photographic
observations in the \.Zabik\'ow PFN09 station.

\begin{figure*}
\centering
\includegraphics{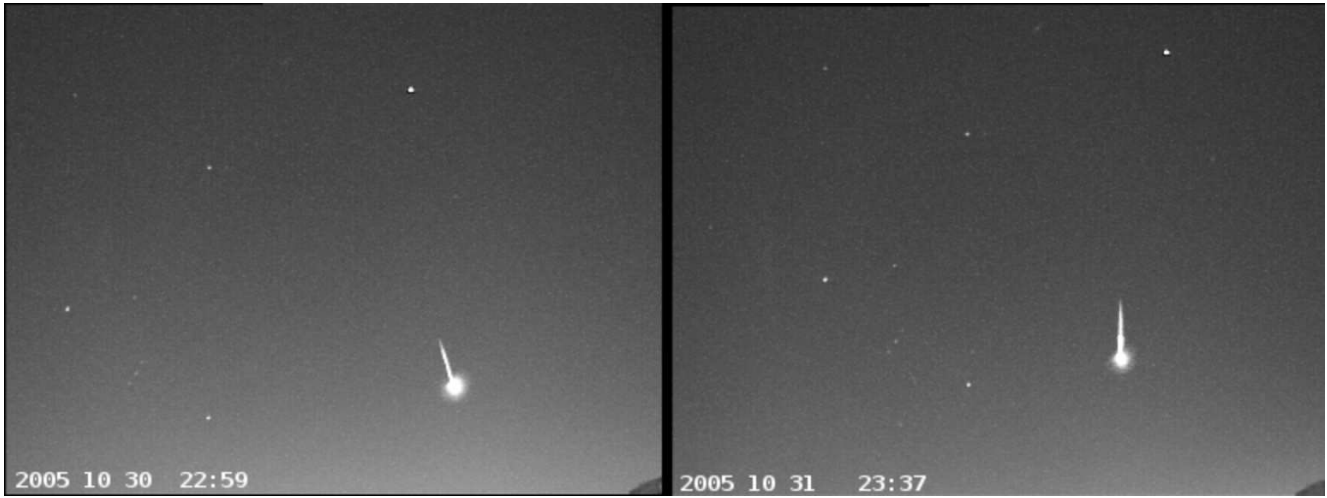}  
\vspace{6.5cm}
\caption{Fireballs from the Southern Taurids shower registered on 30.10.2005 (left) 
and on 31.10.2005 (right) by the PFN05 Pozna\'n station using PAVO5 camera.}
\end{figure*}

\begin{figure}
\centering
\includegraphics{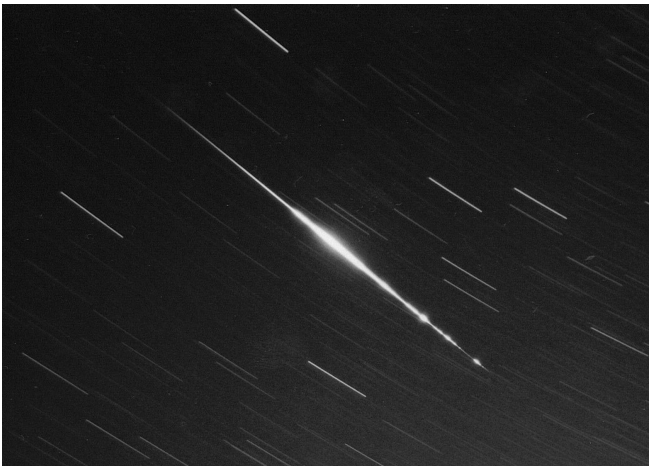}
\vspace{6.6cm}
\caption{Fireball of $-15$ mag registered on 4.11.2005 at 20:19:42 UT with the Pegasus 
constellation in the background. Photo taken by D. Dorosz in PFN09 \.Zabik\'ow station
(Praktica L2, Vivitar 28 mm f/2.5, Konica VX200).}
\end{figure}

A fireball network consisting of just one photographic station and six
video ones was not able to deliver a lot of data. What is more, stations
using early version of MetRec software analyzed data in half PAL
resolution so a huge percentage of double-station phenomena was rejected
because of poor quality astrometry. In the analyzed period we managed to
get from more than two to a dozen or so orbits every night. At the
beginning the majority of observed phenomena were the Orionids with
just a vestigial number of the Taurids.  After 2005 October 27 we
noticed the increase of the Taurids activity with a local peak on
October 31. When it comes to the number of the Taurids as a percentage
of all registered phenomena their peak was observed on November 4. In
the analyzed period we determined orbits of 44 Taurids: 37 Southern and
7 Northern ones. The Northern Taurids constitute just 18\% of the number
of the  Southern Taurids. Overall we determined 107 orbits for shower
and sporadic meteors.

On 2005 October 27 a fireball from the shower of the Southern Taurids
was observed with an apparent magnitude close to $-5$. On October 28 a
similar phenomenon was seen as well. On 2005 October 30, at 22:59 UT
there was a fireball and its brightness in the peak of its flare
exceeded $-8$ mag. The next night we observed a fireball as 
bright as $-6$ mag which belonged to the shower of the Southern Taurids.
The same night we registered the highest number of phenomena belonging
to that shower. On November 4, at 20:19:42 UT, there was visible a
spectacular fireball, as bright as $-15$ mag. It was registered using
the photographic technique by the \.Zabik\'ow PFN09  station. We did not
manage to determine a precise trajectory or orbit of that fireball but
it seems it also belonged to the Southern Taurid shower. We base our
assumption on comparison between eyewitnesses' reports and also the    
photographic observation. Its maximum brightness was observed over the
town of Pu{\l}awy. The last bright fireball, a phenomenon of $-7$ mag,
was observed on 2005 November 6. The fact that in the same period we
registered several very bright phenomena not connected in any way to the
Taurids shower is also worth noticing. Among them there was a sporadic
$-10$ mag fireball, visible in the evening on 2005 October 28 or a
similar event on November 30, with an orbit being an equivalent of the
00241 OUI October Ursae Minorids stream (Molau and Rendtel 2009).

\begin{table*}
\centering
\caption[]{Basic data on the PFN stations which observed Taurid shower maximum in 2015.}
\begin{tabular}{|l|l|c|c|c|l|l|}
\hline
\hline
Code & Site & Longitude [$^\circ$] & Latitude [$^\circ$] & Elev. [m] & Camera & Lens \\
\hline
PFN03 & Z{\l}otok{\l}os & 20.9086 E & 52.0062 N & 128 & Tayama C3102-01A1 & Ernitec 4mm f/1.2 \\
PFN06 & Krak\'ow & 19.9424 E & 50.0216 N & 250 & Mintron 12V6HC-EX & Panasonic 6mm f/0.75 \\
PFN13 & Toru\'n & 18.6209 E & 53.0252 N & 66 & Siemens CCBB1320 & Ernitec 4mm f/1.2 \\
PFN19 & Kobiernice & 19.2018 E & 49.8377 N & 345 & Tayama C3102-01A1 & Panasonic 4.5mm f/0.75 \\
PFN20 & Urz\k{e}d\'ow & 22.1456 E & 50.9947 N & 210 & Tayama C3102-01A1 & Ernitec 4mm f/1.2 \\
PFN32 & Che{\l}m & 23.4982 E & 51.1356 N & 345 & Mintron 12V6HC-EX & Panasonic 6mm f/0.75 \\
PFN37 & Nowe Miasto Lub. & 19.5922 E & 53.4349 N & 95 & Tayama C3102-01A1 & Ernitec 4mm f/1.2 \\
PFN38 & Podg\'orzyn & 15.6817 E & 50.8328 N & 360 & KPF 131 HR & Panasonic 4.5mm f/0.75 \\
PFN40 & Otwock & 21.2494 E & 52.1078 N & 100 & Watec 902 H2	& Computar 3.8mm f/0.75\\
PFN41 & Twardog\'ora & 17.4589 E & 51.3702 N & 178 & Tayama C3102-01A1 & Ernitec 4mm f/1.2 \\
PFN42 & B{\l}onie & 20.6215 E & 52.2147 N & 95 & Watec 902B	& Ernitec 4mm f/1.2 \\
PFN43 & Siedlce & 22.2833 E & 52.2015 N & 152 & Mintron MTV-23X11C & Ernitec 4mm f/1.2 \\
PFN45 & {\L}a\'ncut & 22.2333 E & 50.1039 N & 190 & Tayama C3102-01A1 & Ernitec 4mm f/1.2 \\
PFN46 & Grabniak & 21.7015 E & 51.7131 N & 176 & Tayama C3102-01A1 & Computar 2.9-8.2mm f/1.0 \\
PFN48 & Rzesz\'ow & 21.9220 E & 50.0451 N & 230 & Tayama C3102-01A1	& Computar 4mm f/1.2 \\
PFN49 & Helen\'ow & 22.3316 E & 52.1155 N & 168 & Siemens CCBB1320 & Ernitec 4mm f/1.2 \\
PFN51 & Zel\'ow & 19.2232 E & 51.4698 N & 200 & Tayama C3102-01A1 & Ernitec 4mm f/1.2 \\
PFN52 & Stary Sielc & 21.2923 E & 52.7914 N & 90 & DMK23GX236 & Tamron 2.4-6mm f/1.2 \\
PFN53 & Bel\k{e}cin Nowy & 16.8564 E & 51.8907 N & 114 &Tayama C3102-01A1 & Ernitec 4mm f/1.2 \\
PFN54 & {\L}\k{e}gowo & 18.6430 E & 54.2253 N & 20 & Tayama C3102-01A1 & Ernitec 4mm f/1.2 \\
PFN57 & Krotoszyn & 17.4416 E & 51.7018 N & 150 & Tayama C3102-01A1 & Ernitec 4mm f/1.2 \\
PFN60 & Bystra & 19.1892 E & 49.6215 N & 444 & Mintron 12V6	& Panasonic 6mm f/0.75 \\
PFN61 & Piwnice & 18.5603 E & 53.0950 N & 85 & Tayama C3102-01A1 & Ernitec 4mm f/1.2\\
PFN67 & Nieznaszyn & 18.1848 E & 50.2373 N & 200 & Mintron MTV 23x 11E/1/3 & Panasonic 4.5mm f/0.75 \\
PFN72 & Ko\'zmin Wlkp. & 17.4548 E  & 51.8283 N & 139 & Tayama C3102-01A4 & Lenex 4mm f/1.2 \\
\hline
\hline
\end{tabular}
\end{table*}

\subsection{Observations in 2015}

Like in 2005, our research included solar longitudes ranging from 206 to
228 degrees. This time the Moon phases were not as perfect as in 2005
with Full Moon on October 27 and New Moon on November 11. From October
20 to 22 our stations did not provide any data due to heavy clouds.
Between October 24 and November 4 good weather allowed the PFN to work
at its fullest; nights from November 5 to 7 were partially or completely
cloudy but the nights between November 8 and 9 were clear in most of the
country.

\begin{figure*}
\centering
\includegraphics{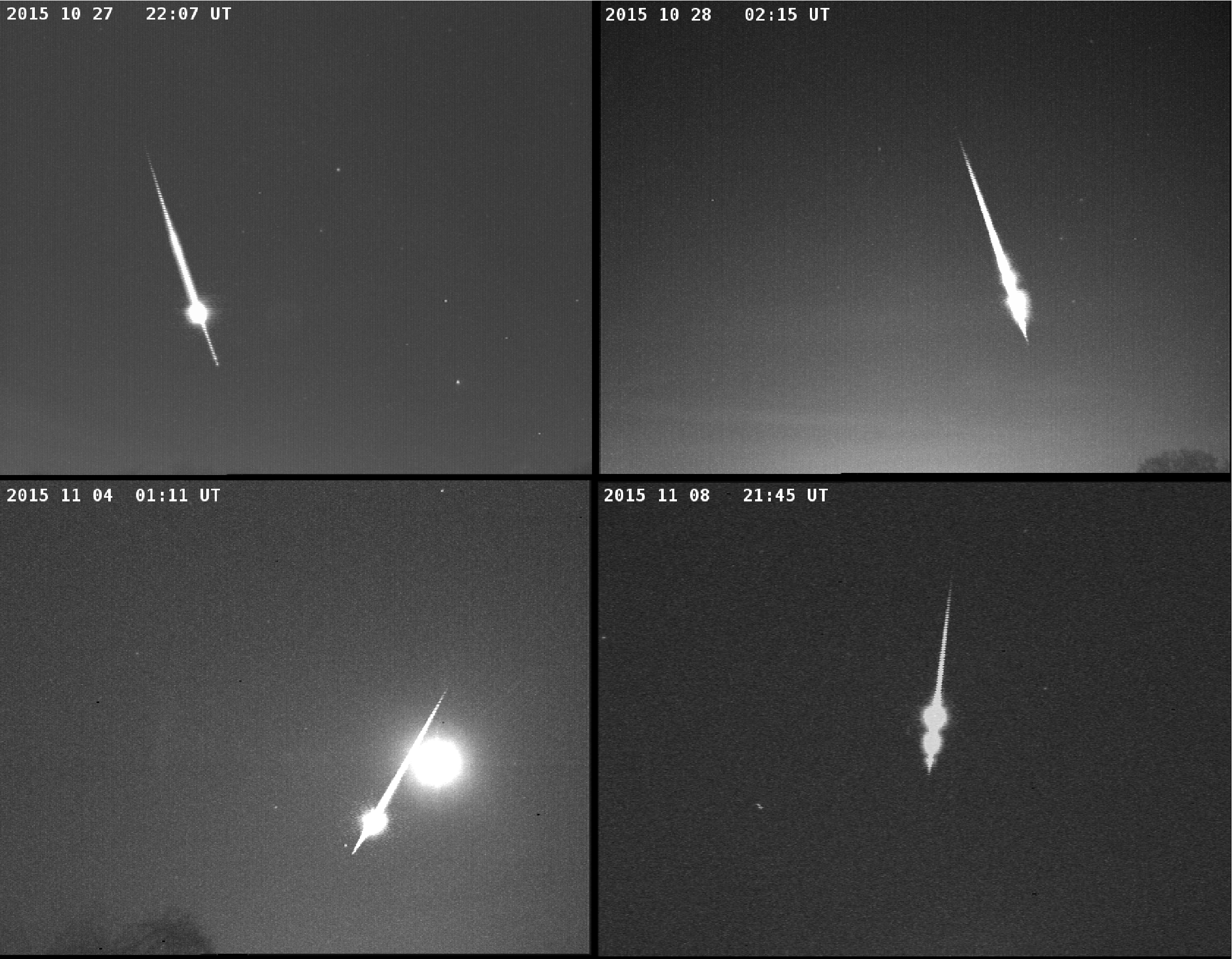} 
\vspace{12.7cm}
\caption{Selection of fireballs registered during the peak of activity of the Taurids in 2015.}
\end{figure*}

Compared to 2005 the number of stations and the quality of our equipment
increased significantly. We were able to gather data from 25 stations
covering a period of time from October 20 to November 10. We used a
quite diverse observation equipment, the sets of the highest speed based
on the Mintrons 12V6 which allowed us to observe meteors as faint as
$3-4$ mag. For the first time we used the digital DMK23GX236  megapixel
camera and the Tamron 2.4-6 mm f/1.2 lens. Such a set allows you to
observe field of $128.3 \times 80.2$ deg with a limiting magnitude of
about $+1 - +2$ and a resolution of $1920 \times 1200$ pixels with 25
frames per second. Table 2 summarizes a list of our stations active in
2015.

In a period from 2015 October 20 to November 10 all PFN cameras
registered 6970 meteors out of which we determined 719 orbits. They
corresponded to different meteor showers and sporadic centers. Out of
that number 215 orbits belonged to the Southern Taurids and 39 to
the Northern Taurids.

The first phenomena of high brightness were observed on October 27
(similarly to 2005). Three bright fireballs were seen on October 30 as
well. The night of October 31 brought two other incredibly bright
phenomena. At 18:05 UT a fireball from the Southern Taurids shower
appeared over the northwestern Poland and its absolute magnitude
amounted to $-16$ mag. The fireball was visible across the Central
Europe and was registered by PFN network cameras, car cameras and
accidental people photographing the close fly-by of the asteroid 2015
TB145 (see Olech et al. 2016 for detailed description). The phenomenon
lasted 5.6 seconds and left a persistent train which was visible for 50
minutes.

A second, just slightly weaker fireball appeared over the western Poland
at 23:13 UT. Its brightness was estimated to $-14.8$ mag and its flight
lasted just 2 seconds. Apart from these two extremely bright phenomena
during the night from October 31 to November 1 four other orbits were
determined for fireballs brighter than $-4$ mag. A fireball of an
absolute magnitude of $-7.3$ mag was the third brightest object which
appeared during that night and was registered at 22:02 UT above the
border between Latvia and Belarus. The next night brought as many as 10
Taurid orbits of objects brighter than $-4$ mag. At 21:13 a fireball of
$-7$ mag appeared on the sky, then at 00:57 UT another fireball of
$-8.5$ mag was observed. The night November 2/3 brought 4 fireballs; the
next night another 8 fireballs including 3 with brightness close to $-6$
mag. In the evening of November 4 the PFN cameras detected 3 fireballs with
one phenomenon as bright as $-7$ mag among them. During the following
nights the weather worsened and it did not improve until November 8.  On
November 8/9 the activity of the Taurid complex was still very high,
with 6 fireballs observed. On November 9, at 01:00 UT, the cameras of
PFN42 and PFN48 stations registered a very distant fireball flying
over the eastern part of Belarus, near Komaje. Its absolute magnitude
reached $-9$ mag. While describing the activity of fireballs you should
emphasize the high brightness of phenomena registered on
October 31/November 01 night and also high number of fireballs observed
during the following nights.

\begin{figure}
\centering
\includegraphics{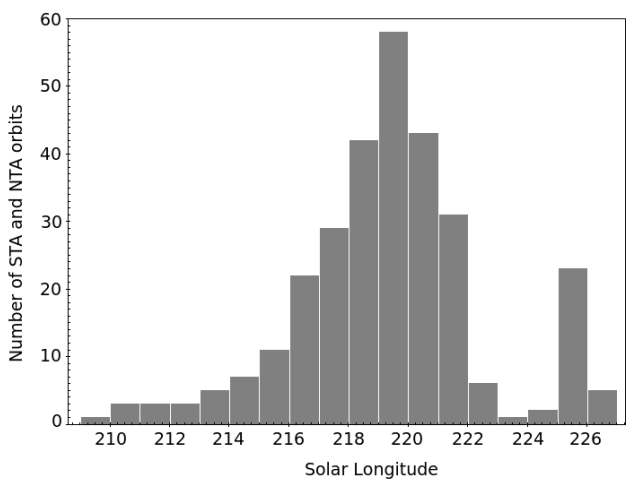}  
\vspace{6.6cm}
\caption{Amount of Southern and Northern Taurid orbits registered during each night of
2015 maximum. A clear peak is visible for solar longitude of 219 deg.}
\end{figure}

Taking into account all events which belonged to the Southern and
Northern Taurids up to October 28 the registered numbers of all the
phenomena from both showers were small and comparable. Starting from
October 28 the Southern Taurids started to dominate. Their peak was
observed on November 2 when 48 orbits were recognized and they
constituted 47\% of all orbits registered that night. During the
following nights there was a decrease of the number of the Taurids
compared to the number of other registered meteors. After a spell of bad
weather the night of November 8/9 was very interesting as the Southern
Taurids constituted 51\% of all registered orbits.

Fig. 4 shows an amount of Southern and Northern Taurid orbits registered
during each night of 2015 maximum with a clear peak which is visible for
$\lambda_\odot=219^\circ$.

\section{Data reduction}

Video data constitute most of information gathered during the Taurids
activity in 2005 and 2015. Most of them are saved in PAL standard. Data
from 2005 were primarily analyzed by MetRec software (Molau 1999). The
first versions of MetRec analyzed images in half PAL resolution; what's
more, in astrometry they used only stars visible directly in the
camera's field of view. 

Later on, better tools were developed so we could use stars from many
detections in order to improve the precision of astrometry. A converter
was also constructed which allows you to analyze data from MetRec with
UFO Analyzer software (SonotaCo 2009). Such tools made it possible to
analyze data from 2005 for the second time, with paying proper attention
to the quality of astrometry. The 2005 data were taken in the form of raw
MetRec output because the contemporary versions of that program use an
appropriate number of stars as points of reference and are able to
generate astrometry of high quality. For several brightest phenomena
astrometric measurements were taken manually with the help of the UFO
Analyzer. That way we were able to eliminate errors due to overexposures
or fragmentation of fireballs. 

Photographic data were measured using AstroRecord software (de Lignie
1997).

Positional data were downloaded to the PyFN software (\.Zo{\l}\k{a}dek
2012). For all nights all meteors constituting double-station or
multi-station events were recognized. Input data for particular
detections were checked in order to eliminate astrometric errors. Points
differing in a distinct way were rejected, events with grave astrometric
mistakes were eliminated from calculations. Using the PyFN software the
trajectories and orbits were determined for all meteors registered
between October 20 and November 10 of 2005 and 2015. The orbits were
classified and assigned to particular meteor showers by comparing them
to the average orbit of a given stream taken from IAU Meteor Data 
Center\footnote{https://www.ta3.sk/IAUC22DB/MDC2007/}. Phenomena which belonged to
Southern and Northern Taurids were selected. Orbital similarities were
researched for that group of objects using the NEODYS-2 database with
Drummond criterion - $D'$ (Drummond 1981). Initially we selected
asteroids for which $D'<0.105$.

Photometry of registered phenomena was based on automatic measurement
from the MetRec program. Because of overexposures and a different
characteristic of converters used by particular cameras those brighter
areas come with errors of 1 mag. For the brightest events the photometry
was conducted manually, with such comparison objects as the Moon and
planets registered in the same settings by the equipment of the same
parameters. The details of this procedure are given in Olech et al.
(2016).

\begin{table*}
\centering
\caption[]{Basic parameters of the brightest fireballs of Taurid shower maxima
in 2005 and 2015.}
\begin{tabular}{|l|c|c|c|c|c|c|c|c|l|}
\hline
\hline
Fireball & Brightness & $V_{\rm geo}$ & $\Delta V_{\rm geo}$ & $h_{\rm beg}$ & $\Delta h_{\rm beg}$ & 
$h_{\rm end}$ & $\Delta h_{\rm end}$ & Shower & Asteroid ($D'$) \\  
 & [mag] & [km/s] & [km/s] & [km] & [km] & [km] & [km] &  &  \\

\hline
20051027PFN033235 & -5 & 31.4   & 0.2 & 103.21  & 0.07 & 71.13 & 0.02 & STA & 2005 UR (0.0198)\\
20051028PFN203838 & -4 & 31.1   & 0.05 & 101.80  & 0.04 & 64.14 & 0.03 & STA & 2015 TX24 (0.0173)\\
20051030PFN225900 & -8 & 30.1   & 0.08 &  97.66  & 0.04 & 67.72 & 0.04 & STA & 2015 TX24 (0.0403)\\ 
20051031PFN233748 & -6 & 29.9   & 0.07 &  96.83  & 0.05 & 70.03 & 0.04 & STA & 2015 TX24 (0.0366)\\   
20051106PFN224853 & -7 & 29.7   & 0.11 & 101.84  & 0.04 & 75.96 & 0.06 & STA & 2010 TU149 (0.0466)\\
\hline
20151027PFN220747 & -7 & 31.7   & 0.09 & 101.43  & 0.13 & 60.15 & 0.11 & STA & 2005 UR (0.0204)\\    
20151028PFN021507 & -6 & 32.3   & 0.17 & 101.08  & 0.07 & 58.22 & 0.16 & STA & 2005 UR (0.0342)\\  
20151030PFN014256 & -4 & 29.9   & 0.51 &  97.34  & 0.44 & 62.18 & 0.35 & NTA & 2004 TG10 (0.0556)\\
20151030PFN225233 & -6 & 31.9   & 0.22 &  99.22  & 0.11 & 76.64 & 0.07 & STA & 2015 TX24 (0.0254)\\  
20151030PFN232033 & -4 & 30.7   & 0.15 & 102.63  & 0.10 & 59.70 & 0.20 & STA & 2015 TX24 (0.0313)\\
20151031PFN020341 & -6 & 30.3   & 0.14 &  95.53  & 0.08 & 64.53 & 0.10 & STA & 2015 TX24 (0.0172)\\
20151031PFN180514 & -16 & 31.0   & 0.10 & 117.88  & 0.05 & 60.20 & 0.20 & STA & 2015 TX24 (0.0144)\\
20151031PFN193846 & -5 & 30.6   & 0.17 & 104.38  & 0.55 & 65.15 & 0.15 & STA & 2015 TX24 (0.0315)\\ 
20151031PFN200534 & -4 & 30.4   & 0.11 & 99.78   & 0.06 & 66.11 & 0.04 & STA & 2015 TX24 (0.0404)\\
20151031PFN231258  & -15 & 31.2  & 0.11 & 108.05  & 0.02 & 57.86 & 0.03 & STA & 2015 TX24 (0.0055)\\
20151101PFN200917 & -4 & 30.6   & 0.10 & 99.99   & 0.20 & 58.20 & 0.13 & STA & 2015 TX24 (0.0389)\\
20151101PFN211329 & -7 & 30.9   & 0.47 & 91.12$^*$  & 0.64 & 61.91 & 0.26 & STA & 2007 RU17 (0.0471)\\  
20151101PFN212945 & -5 & 29.9   & 0.68 & 103.71  & 0.50 & 62.10 & 0.29 & STA & 2015 TX24 (0.0543)\\
20151101PFN223333 & -4 & 30.6   & 0.10 &  98.68  & 0.08 & 58.43 & 0.03 & STA & 2015 TX24 (0.0219)\\
20151101PFN223725 & -5 & 29.6   & 0.10 &  95.55  & 0.08 & 63.24 & 0.07 & STA & 2003 UV11 (0.0564)\\ 
20151101PFN233130 & -4 & 30.1   & 0.26 &  92.72  & 0.18 & 67.91 & 0.38 & STA & 2015 TX24 (0.0202)\\   
20151101PFN234423 & -6 & 30.5   & 0.13 &  94.49  & 0.05 & 64.62 & 0.07 & STA & 2015 TX24 (0.0408)\\ 
20151102PFN005736 & -8 & 29.7   & 0.08 &  98.47  & 0.08 & 61.32 & 0.09 & STA & 2015 TX24 (0.0467)\\
20151102PFN020745 & -4 & 30.5   & 0.42 &  88.89$^{**}$ & 0.2 & 62.97 & 0.67 & STA & 2015 TX24 (0.0560)\\  
20151102PFN020948 & -5 & 30.5   & 0.21 &  96.98  & 0.20 & 67.75 & 0.19  & STA & 2015 TX24 (0.0204)\\
20151102PFN233754 & -4 & 35.2   & 0.18 & 102.49  & 0.11 & 73.83 & 0.13 & STA & 2001 VB (0.0781)\\  
20151103PFN012403 & -5 & 29.3   & 0.11 &  98.94  & 0.12 & 60.18 & 0.06 & STA & 2003 UV11 (0.0552)\\  
20151103PFN195652 & -6 & 29.8   & 0.22 & 100.93  & 0.27 & 68.69 & 0.48 & STA & 2003 UV11 (0.0605)\\
20151103PFN212017 & -5 & 29.3   & 0.11 &  97.36  & 0.13 & 48.27 & 0.05 & STA & 2007 RU17 (0.0477)\\
20151103PFN230449 & -4 & 28.7   & 3.18 &  96.5   & 2.2 & 60.8  & 2.0 & STA & 2007 RU17 (0.0468)\\   
20151104PFN011153 & -6 & 30.4   & 0.13 & 101.22  & 0.08 & 58.29 & 0.08 & STA & 2015 TX24 (0.0378)\\ 
20151104PFN022941 & -6 & 30.1   & 0.11 & 102.42  & 0.12 & 63.16 & 0.08 & STA & 2015 TX24 (0.0494)\\
20151104PFN025955 & -5 & 30.0   & 0.21 & 102.07  & 0.52 & 68.33$^*$ & 0.48 & STA & 2007 RU17 (0.0679)\\ 
20151104PFN034420 & -5 & 29.3   & 0.26 & 96.43   & 0.22 & 66.26 & 0.19 & STA & 2003 UV11 (0.0505)\\
20151104PFN040817 & -4 & 31.6   & 0.67 & 91.62   & 0.55 & 68.42 & 2.2 & STA & 2010 TU149 (0.0733)\\  
20151104PFN190358 & -4 & 28.6   & 0.15 & 91.29   & 0.19 & 74.67 & 0.13 & STA & 2003 UV11 (0.0391)\\
20151104PFN203853 & -7 & 29.6   & 0.12 & 100.61 & 0.11 & 67.48 & 0.11 & STA & 2003 UV11 (0.0618)\\ 
20151104PFN233903 & -4 & 29.2   & 1.48 & 91.20 & 0.70 & 73.78 & 0.44 & STA & 2003 UV11 (0.0555)\\   
20151106PFN005011 & -5 & 28.1   & 0.57 & 98.30   & 0.54 & 66.10 & 0.71 & STA & 2007 UL12 (0.0393)\\   
20151108PFN180642 & -4 & 27.4   & 0.55 & 101.1   & 1.8 & 77.35 & 0.19 & STA & 2010 TU149 (0.0363)\\ 
20151108PFN201906 & -5 & 28.4   & 0.32 & 96.36   & 0.52 & 68.41 & 0.21 & STA & 2007 UL12 (0.0316)\\
20151108PFN210603 & -4 & 28.8   & 0.16 & 94.86 & 0.16 & 76.23 & 0.07 & STA & 2010 TU149 (0.0293)\\   
20151108PFN214524 & -6 & 27.8   & 0.94 & 99.95 & 1.9 & 68.94 & 0.57 & STA & 2007 UL12 (0.0271)\\
20151108PFN234417 & -6 & 27.5   & 0.37 & 85.69$^{**}$ & 0.16 & 64.13 & 0.15 & STA & 2007 UL12 (0.0222)\\
\hline
\multicolumn{10}{l}{$^*$ - trajectory incomplete, $^{**}$ - altitude value 
lowered due to great distance from the event}\\
\hline
\hline
\end{tabular}
\end{table*}

\section{Results}

\subsection{The brightest phenomena}

Both peaks of Taurids activity, in 2005 and in 2015, were full of very
bright phenomena. For 2005 data there were 5 evens with magnitude over
$-4$ mag out of 44 for which orbits and trajectories were determined. It
constitutes 11\% of all phenomena. In 2015 the number of events brighter
than $-4$ mag was 43 out of 243 determined orbits which constitutes 17\%
of all phenomena. Both peaks were characterized by occurrences of
extremely bright events. In 2005 a fireball of $-15$ mag from the Taurid
shower was observed but, unfortunately it was detected by only one
station. During the 2015 peak two fireballs with a magnitude of $-16$
and $-14.7$ were registered. In the case of both peaks the first very
bright events appeared for $\lambda_\odot=214^\circ$ (around October
27), in 2015 very bright fireballs were observed at
$\lambda_\odot=218^\circ$ and during the following nights the number of
bright evens remained significant, with a strong emphasis on the evening
of 2015 November 8 ($\lambda_\odot=226^\circ$) when, averagely, every
hour a very bright phenomenon was observed. Table 3 presents basic
parameters of double-station fireballs observed both in 2005 and 2015
with its errors. Almost all phenomena brighter than $-4$ mag belong to
the Southern Taurid shower and only one observed on 2015 October 30
belongs to the Northern Taurids. 

A downward trend of geocentric velocity is clearly noticeable when it
comes to observed fireballs - at the beginning of their activity a
typical geocentric velocity exceeded 31 km/s but for the night of
November 8 it was closer to 28 km/s. The starting and final altitudes of
observed fireballs are also quite interesting. As the velocities were
very similar to each other in the examined sample the dimension of the
body entering the atmosphere should be a decisive factor, influencing
the altitudes. Of course the brightest fireballs had initial altitudes
higher than average, for the fireball visible on 2015 October 31 at
18:05 UT it was 117 km and for the fireball at 23:12 UT it was 108 km.
In the case of the majority of fireballs observed during both peaks the
initial altitude ranges from 97 to 103 km. The final altitude does not
depend on brightness. When it comes to the brightest October 31
fireballs the final altitudes were, respectively, 62 and 58 km which
does not differ from the rest of events, presented in the table. The
phenomena which end with a flare usually have the final altitudes
exceeding 70 km. The $-5$ mag fireball seen on 2015 November 3  was a
special case - its initial altitude was 97 km and final altitude only 48
km. When it comes to orbital parameters and the entry velocity that
fireball does not differ in any way from other Taurids. Most likely that
phenomenon was triggered by a meteoroid with a smaller ablation tendency
which suggests in turn that the Taurids shower can contain also bodies
with higher density and resistance (Madiedo et al. 2014).

The errors in geocentric velocity and trajectory determination shown in
Table 3 result in errors in orbital parameters determination. It is
worth to note that typical errors for semimajor axis $a$, perihelion
distance $q$, eccentricity $e$ and inclination $i$ in observed fireballs
are around 0.01 AU, 0.001 AU, 0.002 and 0.1 deg, respectively. In case
of argument of periapsis $\omega$ the error is around 0.3 deg, and for
longitude of the ascending node $\Omega$ it is lower than $10^{-7}$ deg.
The errors in orbital elements of NEOs discussed in the text are typically one
order of magnitude lower than errors for fireballs.

Similarities between fireball orbits and all asteroid orbits from the
NEODYS-2 catalogue were examined (in fact the PyFN software compares
orbits of fireballs not only to NEO orbits but also to the orbits of
comets from JPL and meteoroids recorded by other fireball networks). The
last column of the Table 3 includes marking of the body which similarity
in orbital parameters to the studied fireball is the greatest. Fireballs
from the very beginning of the activity period are like the asteroid
2005 UR (and, in the second order, like the orbit of the 2015 TX24).
Fireballs registered at the end of October and the beginning of November
had orbits very similar to 2015 TX24. In the case of the 2015 October
31, 23:12 UT fireball the similarity to that asteroid amounts to
$D'=0.0055$ only and for the 18:05 UT phenomenon $D'=0.0144$. Fireballs
with orbits very similar to the orbit of 2015 TX24 were observed between
October 28 and  November 2 both in 2005 and in 2015. Between November 3
and 6 the similarity to orbits of the  2003 UV11 and the  2007 RU17
becomes dominant; still it is not as clear as in the case of the 2015
TX24 and typical $D'$ values are around 0.05. The night of November 8/9
was dominated by bright Taurids with orbits similar to orbits of the
2007 UL and the 2010 TU149. In that case the orbits look very much the
same, with low $D'$ values ranging from 0.02 to 0.03. 

Below you can find a short description of the most interesting phenomena.

\begin{itemize}

\item {\bf 2005 October 30, 22:59 UT and 2005 October 31, 23:38 UT}.
Two similar fireballs with brightness of $-8$ and $-6$ mag. In both
cases the maximum brightness was reached in the final flare which was
connected with a disintegration of the meteoroid. The beginning altitudes
are the same and equal to 97 km, the terminal points are at the heights of
70 and 67 km, respectively. Both fireball were observed near Opole and
both are similar in orbital elements to the asteroid 2015 TX24.

\item {\bf 2005 November 04, 20:20 UT}. 
The fireball was registered by only one PFN station using the
photographic method so there are no precise data concerning the
trajectory and orbit. The photo and visual observations allowed us to
confirm that the fireball belonged to the Southern Taurids shower. The
phenomenon's magnitude was $-15$ and it was the brightest fireball
observed during the 2005 peak. The light curve was typical for very
bright Taurids, with a distinct maximum and numerous flares in the final
part. After the flight of the fireball the persistent train lasted
several minutes. The trajectory was most likely over the town of
Pu{\l}awy, about 150 km south from Warsaw. 

\item {\bf 2005 November 11, 22:48 UT}.
A fireball of $-7$ mag observed over northern Poland. That brightness was
reached during one of two strong flares at the end of its trajectory;
for most of trajectory the magnitude of the fireball did not exceed $-2$
mag. The end of trajectory was at 76 km and the orbit is similar to the orbit
of the 2010 TU149.

\item {\bf 2015 October 27, 22:07 UT}.
A fireball of a magnitude of $-7$ mag in its flare. Observed over
southwestern Poland. Its initial altitude was 101 km; then there was a
single flare at 70 km, after that a small part continued the flight to
an altitude of 60 km. Orbital similarity to the 2005 UR.

\item {\bf 2015 October 28, 02:15 UT}. 
A $-6$ mag fireball with a relatively flat light curve and with $h_{beg}
= 101$, $h_{end} = 58$ km. Brightening visible in the final part Orbital
similarity to the 2005 UR.

\item {\bf 2015 October 31, 18:05 UT}.
An exceptionally bright fireball from the Southern Taurids swarm
registered soon after the sunset. The phenomenon reached
its highest magnitude of $-16$. The trajectory was located over
northwestern Poland, it had a length of 181 km and lasted 5.6 seconds.
The light curve in the initial part of trajectory was flat, after the
peak there were many flares and brightenings. The persistent train was visible for
about 50 min (practically until the Moon rise). The peak of
brightness over a town called Okonek. Orbital similarity to the
2015 TX24 (for more details see Olech et al. 2016).

\item {\bf 2015 October 31, 22:03 UT}.
A very distant detection of a fireball from the Southern Taurids. Its
absolute brightness was over $-7$ mag, the trajectory was located over the
border between Latvia and Belarus and the peak of its brightness near
Borysov in Belarus. The radiant of the phenomenon was in accordance with
the radiant of the Southern Taurids but, due to distinct velocity
errors, we did not use it in further analysis.

\item {\bf 2015 October 31, 23:12 UT}.
A fireball of $-14$ mag registered over Ostrowite in northern Poland.
The event was characterized by a steep trajectory with inclination
amounting to 53 degrees and lasted 2 seconds. The maximum brightness was
reached in a strong, distinct peak. PFN43 Siedlce station managed to get a
low-resolution spectrum of the fireball. The orbit is almost identical
with the orbit of the asteroid 2015 TX24 with $D'=0.0055$ (for more
details see Olech et al. 2016).

\item {\bf 2015 November 02, 00:57 UT}.
A $-8$ mag fireball registered over central Poland with a flare ending the
trajectory at 61 km. The beginning of the fireball was observed at 98 km. Orbital
similarity with the 2015 TX24.

\item {\bf 2015 November 03, 21:20 UT}.
A fireball of $-5$ magnitude registered about 40 km south-west from the
city of Lublin. A phenomenon of a length of 65 km and an atypically flat
light curve, completely devoid of any flares, with a gradual brightness
decrease in the final part. Its initial altitude amounted to 97 km and
final altitude was far from typical values of the rest of the Taurids
and amounted to just 48 km. Very distinct braking visible on the
velocity curve, the final speed almost twice slower than the speed at
the beginning (about 16 km/s). Despite quite different traits the orbit
of the fireball was typical for the Southern Taurids and similar to the
orbit of the 2007 RU17 (other fireballs with orbits similar to
the 2007 RU17 have not had such properties).

\item {\bf 2015 November 04, 01:11 UT}.
A fireball with an initial altitude of 101 km and a strong flash at 65
km. The flash did not cause complete disintegration; the fireball was
visible, its brightness almost unchanged, till 58 km. The brightness of
the flash exceeded $-6$ mag, for the rest of trajectory the magnitude
reached $-3$ mag. Orbital similarity to the 2015 TX24. 

\item {\bf 2015 November 04, 20:38 UT}.
A $-7$ mag fireball which trajectory ran near Z{\l}oczew. The light
curve with three noticeable brightenings and small flares near the end
($h_{beg} = 101$, $h_{end} = 67$ km). Orbital similarity to the 2003 UV11.

\item {\bf 2015 November 08, 21:45 UT}.
One of the brightest phenomena that night. A $-6$ mag fireball with
three flares, registered over northern Poland with $h_{beg}
= 100$ and $h_{end} = 69$ km. Similarity to the 2007 UL and the 2010 TU149
asteroids.

\item {\bf 2015 November 09, 01:00 UT}.
A fireball observed by two PFN cameras but only from a great distance,
the brightest event of that night. The trajectory over Komaje in eastern
Belarus. Absolute magnitude of $-9$ mag, the radiant overlapping with
the radiant of the Southern Taurids. Due to the significant distance
there were significant velocity errors so the phenomenon was not used in
further analysis. 

\end{itemize}

\subsection{Determining a threshold while studying the similarity of orbits}

The meteoroid orbits which belonged to the Southern and Northern Taurids
were initially compared to orbits of NEO objects from the NEODYS-2
database. The comparison was made with the help of the Drummond
criterion. It turned out that, within the limits of the usual threshold
$D'<0.105$ orginally suggested by Drummond (1981), for every meteoroid
orbit there were many comparable objects. Few of them were situated on
very similar orbits; for most of them the probability value was close to
the threshold. There was a risk that, with such a database as the
NEODYS-2, including so many objects, with orbits compared to those
situated in the ecliptic plane a coincidental similarity of orbits is
possible. There was a need to determine a new threshold for the NEODYS-2
base and orbits of such type. 

In order to do that 360 clones were created, with their orbits based on
those belonging to the typical Southern Taurids. The orbit of reference
was an average Southern Taurids orbit from the IAU MDC database
according to Jenniskens et al. (2016). The clones were modified in the
orbital node $\Omega$ as its value was changed gradually by one degree
and so an even distribution of the value of the node was reached (the
argument of perihelion $\omega$ was kept fixed in this procedure). The
resulting orbits had shapes and dimensions identical with those
characteristic of the Taurids; also their inclination to the plane of
ecliptic was identical but they were moved in the node every one degree,
creating a ring of evenly spread orbits. Clones defined in such a way
were compared to the NEODYS database. For every one of them there were
bodies found which met the $D'<0.105$ criterion. Figure 5 presents
similarity values for all 360 clones. The X axis shows the value of the
ascending node for the simulated object, the Y axis includes orbital
probability for all bodies which meet the $D'<0.105$ condition. Points
of similarities create local maxima where the similarity between a NEO
asteroid and a clone is the most obvious.

\begin{figure}
\centering
\includegraphics{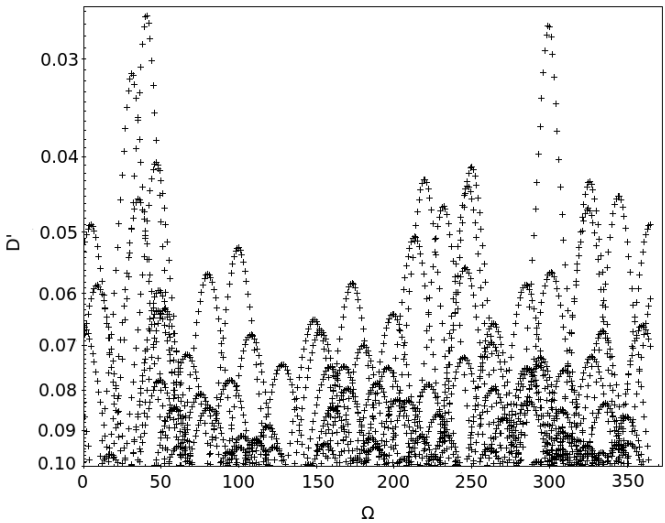}
\vspace{6.9cm}
\caption{Values of Drummond criterion similarities to the NEODYS-2 catalogue bodies for 
meteoroid clones which were based on average orbit of the Southern Taurids, modified in
the ascending node.} 
\end{figure}

The graph at Fig. 5 shows many local maxima which represent similarities
of the clones to different NEODYS-2 database objects. Most of the
similarities create a quite unanimous background which meets the
$D'>0.06$ criterion in its majority. Below the 0.06 value you find just
few asteroids or small groups of them. The highest peak, visible at 40
degrees is an equivalent of the longitude of the Southern Taurids node.
That peak is created by several asteroids. The biggest similarity in
that group is visible for the 2003 UV11, 2007 RU17, 2007 UL12 and 2010
TU149 asteroids. The similarity to these objects had been found earlier
for fireballs observed after November 3, 2005 and 2015. At 299 degrees
there is a single peak visible connected to the asteroid 2013 NE19. We
did not manage to connect that object with any big stream of meteors.
Also we did not found any corresponding meteor showers for other bodies
which similarity $D'$ is lower than 0.06. The presence of a distinct
maximum, created by several asteroids for the ecliptic longitude close
to 40 degrees, is the proof that there is a group of asteroids on orbits
very similar to the orbit of the Southern Taurids. The probability of
accidental coincidence is minimal.  Following the comparison of Taurid
orbits and NEO objects, presented above, it seems that $D'<0.06$
threshold eliminates successfully similarities with accidental NEO
objects.

\begin{table}
\centering
\caption[]{The NEODYS-2 catalogue objects with $D'<0.06$ similarity to the Southern Taurids.
For each object its absolute magnitude $H_0$ and size $D$ are given. The sizes are computed
from $H_0$ or taken from Mainzer et al. (2016).}
\begin{tabular}{|l|c|c|c|c|}
\hline
\hline
Asteroid & $H_0$ & Size & Number of & Lowest \\
         & [mag] & [m]  & meteors & $D'$ \\
\hline
2015 TX24  & 21.5 & $151-337$ & 102 & 0.0055\\
2005 UR    & 21.6 & $144-322$ &  51 & 0.0158\\
2007 UL12  & 20.9 & $199-445$ &  59 & 0.0222\\
2010 TU149 & 20.7 & $603$     &  86 & 0.0222\\
2005 TB15  & 19.5 & $379-848$ &  15 & 0.0236\\
2003 UV11  & 19.5 & $260$     &  97 & 0.0249\\
2007 RU17  & 18.1 & $722-1616$ & 70 & 0.0296\\
2005 TF50  & 20.3 & $262-586$  & 86 & 0.0407\\
2011 UD    & 20.7 & $218-488$  & 10 & 0.0432\\
1999 VK12  & 23.7 & $54-122$ & 1 & 0.0524\\
\hline
\hline
\end{tabular}
\end{table}

\begin{table*}
\centering 
\caption[]{Orbital parameters of selected asteroids taken from NEODYS catalogue
and sorted according to semi-major axis.}
\begin{tabular}{|l|c|c|c|c|c|c|c|}
\hline
\hline
Asteroid & $a$ [AU] & $q$ [AU] & $e$ & $i$ & $\omega$ & $\Omega$ & $P$ [days] \\
\hline
2005 TF50 & 2.272 & 0.2978 & 0.8689 & 10.69 & 159.9 & 0.664 & 1250\\
2015 TX24 & 2.2688 & 0.2896 & 0.8723 & 6.044 & 127.01 & 33.007 & 1248\\
2005 UR & 2.262 & 0.2723 & 0.8796 & 6.972 & 141.03 & 19.553 & 1243\\
1999 VK12 & 2.2358 & 0.4998 & 0.7764 & 9.504 & 103.06 & 48.635 & 1221\\
2010 TU149 & 2.2016 & 0.3778 & 0.8283 & 1.972 & 91.712 & 59.717 & 1193\\
2007 RU17 & 2.0400 & 0.3507 & 0.8281 & 9.081 & 129.83 & 17.469 & 1064\\
2011 UD & 2.0324 & 0.4454 & 0.7808 & 8.823 & 146.81 & 357.31 & 1058\\
2007 UL12 & 1.9666 & 0.3815 & 0.806 & 4.185 & 95.626 & 67.148 & 1007\\
2005 TB15 & 1.8122 & 0.4425 & 0.7558 & 7.291 & 139.08 & 9.549 & 891\\
2003 UV11 & 1.4527 & 0.3444 & 0.7629 & 5.928 & 124.79 & 31.931 & 639\\
\hline
\hline
\end{tabular}
\end{table*}

\subsection{Similarity between asteroid and observed meteors}

In Table 3 we presented orbital similarity for events brighter than $-4$
mag. In the next step we researched orbital similarity for all Southern
Taurids, registered in the given period of time. Only cases of the
Southern Taurids meeting the $D'<0.06$ criterion when compared to a NEO
object were taken into account. Table 4 presents a list of all
asteroids which were found similar to the studied orbits. The
asteroids were sorted according to their maximum orbit similarity.

Definitely the biggest similarity was found for the 2015 TX24 asteroid.
It is an object similar to many bright fireballs from 2005 and 2015,
including two brightest fireball of October 31, 2015. It is also an object for
which most of meteoroid orbits meet the $D'<0.06$ condition. The
following places are occupied by the 2005 UR, the 2007 UL12 and the 2010
TU149 asteroids. The asteroid 2005 TV15 is also worth your attention. We found
only 15 Taurids which $D'$ was lower than 0.06  for this object but
most of Taurids in this group are similar to the 2005 TB15 on a level of
0.02-0.04.

Sorting the NEO objects in semi-major axis (and, what follows, in the
orbital period) you can notice a group with a semi-major axis close to
2.25 AU and the orbital periods close to 1240 days (see Table 5). The
2015 TX24 and the 2005 UR asteroids are situated on almost identical
orbits; the 2005 TF50 has a distinctly higher orbit inclination but the
perihelion distance, semi-major axis and orbital period are also very
similar. The 2010 TU149 with a similar semi-major axis has a
noticeably higher perihelion value. The orbital periods of the objects,
mentioned here, are close to a period typical for the average 7:2
resonance with Jupiter which lasts 1238 days.

The orbit of the 2005 TB15 has a much smaller semi-major axis,
with the orbital period of 891 days. The object is characterized by a
quite high perihelion distance and the smallest eccentricity. The 2003
UV11 which was found similar to as many as 97 Southern Taurid
orbits has an orbit distinctively different than the others with
semi-major axis of 1.45 AU and the orbital period which is equal to only
639 days.

\begin{figure}
\centering
\includegraphics{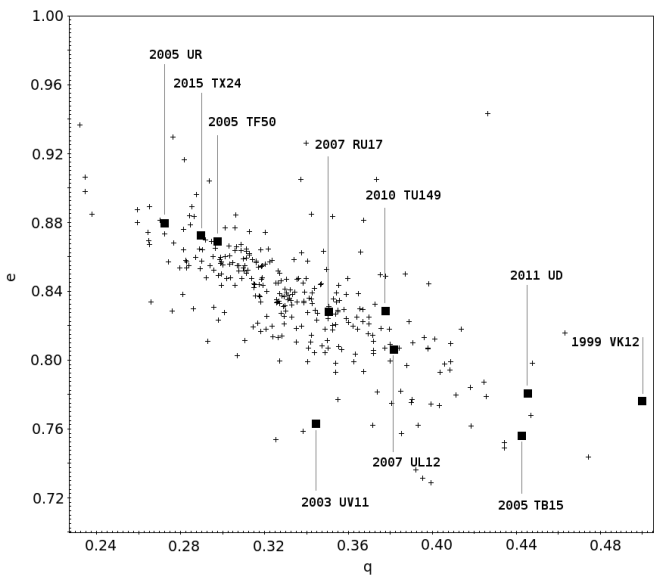} 
\vspace{7.8cm}
\caption{Perihelion and eccentricity of orbits of studied Taurids (marked
by crosses) and NEOs (marked with black squares).}
\end{figure}

\begin{figure}
\centering
\includegraphics{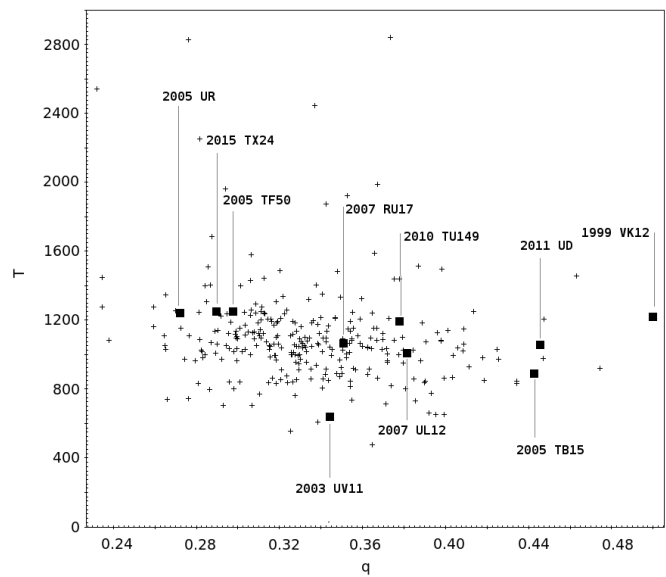} 
\vspace{7.8cm}
\caption{Perihelion distance and orbital periods of studied Taurids (marked
by crosses) and NEOs (marked with black squares).}
\end{figure}

\begin{figure*}
\centering
\includegraphics{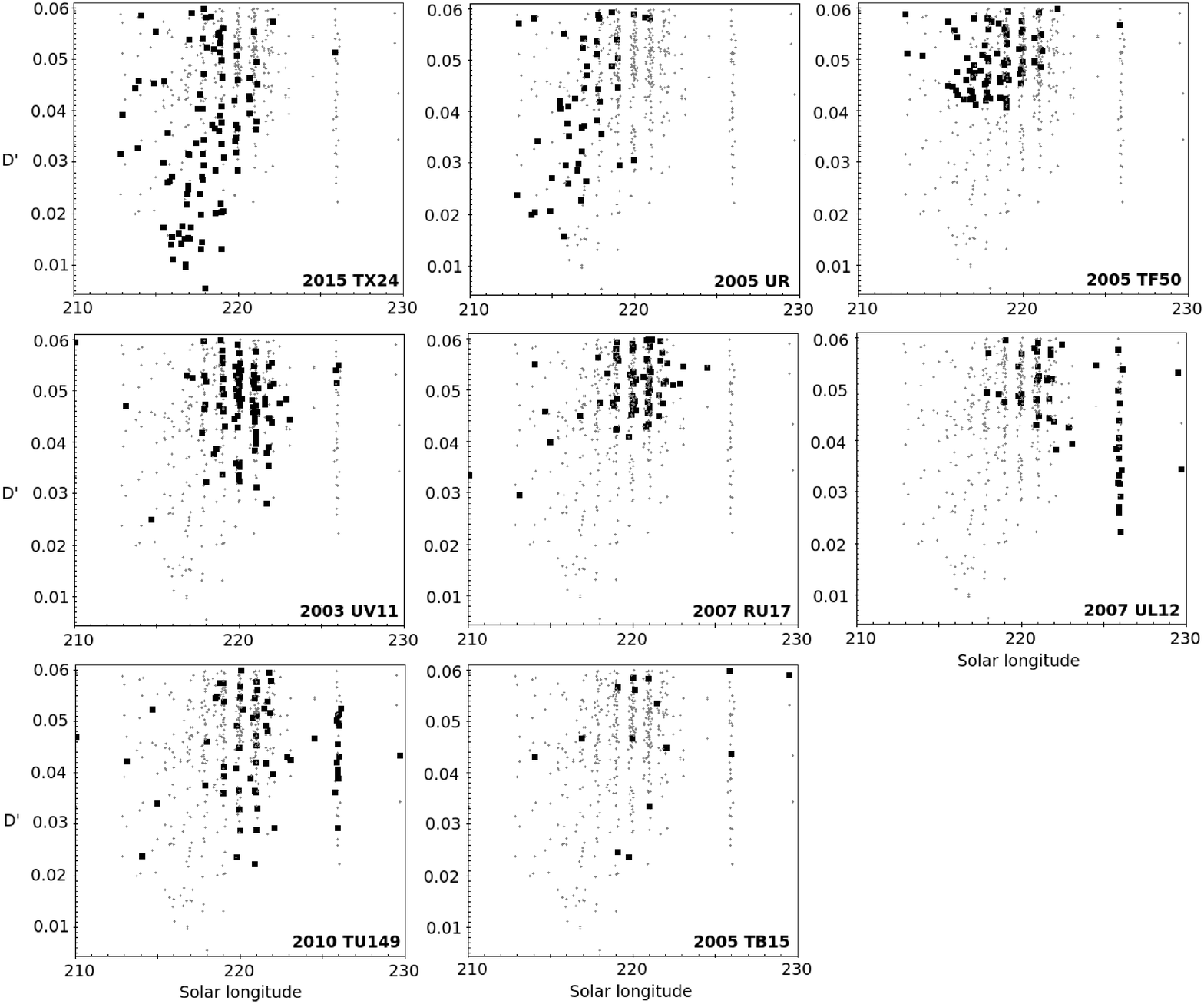}
\vspace{15cm}
\caption{Similarities between the observed Taurids (crosses) and NEO objects (squares) in function
of solar longitude.}
\end{figure*}

Fig. 6 shows the eccentricity-perihelion value graph for all Taurids
for which precise orbital elements were determined and for NEOs from
Table 5.

The majority of registered Taurids have perihelions ranging from 0.26 to
0.44 AU. The eccentricity of the orbit is between 0.44 and 0.88 and you
can notice a clear relation between the perihelion distance and the
eccentricity. The same relation is, approximately, seen for most
asteroids with orbits similar to orbits of the Taurids. Fig. 6 shows
that objects similar to the brightest Taurids of October 31, 2015 concur
with the NEOs with the smallest perihelion distance and highest
eccentricity - these are the 2005 UR, the 2015 TX24 and the 2005 TF50. 
The second group is found for perihelion distances from 0.34 to 0.38 and
it includes such asteroids as the 2007 RU17, the 2010 TU149 and the 2007
UL12.

Among few Taurids having the perihelion distance higher than 0.4 AU we
can find the 2015 TB15 and the 2011 UD asteroids. You should notice the
position of the asteroid 2003 UV11 on the graph. It is an object very
similar to many Taurids but the orbital elements of the observed Taurids
differ from orbital elements of that asteroid. 

Fig. 7 shows perihelia and orbital periods of observed Taurids and NEO
objects. Asteroids like the 2005 UR, the 2015 TX24, the  2005 TF50 and
the 2010 TU149 are the nearest of 7:2 resonance with Jupiter. A bit
shorter orbital periods are found for the 2007 RU17, the 2007 UL12 and
the 2005 TB15 asteroids. Also in this case we think the 2003 UV
is situated on the orbit with a shorter orbital period. In this place
you can find a small number of Taurids with a great orbital resemblance
to the 2003 UV11. Most of the objects, with the perihelion distance
characteristic for the 2003 UV11 feature much longer orbital periods.
Orbital periods of most of observed Taurids range from 1000 to 1300
days, with a noticeable cut-off over 1300 days. Among observed Taurids
there is a quite homogeneous distribution of orbital elements which do not
create any noticeable groups. The NEO objects orbiting inside the Taurids'
stream do not seem to be responsible for the existence of the current
activity; they can be considered as the biggest among objects which
belong to the stream of the Southern Taurids.

Fig. 8 shows similarities between observed Taurids and NEO objects in
the function of solar longitude ($\lambda_\odot$). Taurids observed
around $\lambda_\odot=215^\circ$ seem to be most similar to the 2005 UR.
The similarity is significant here, reaching $D'=0.015$.   At
$\lambda_\odot=218^\circ$ you can observe a very distinct peak due to
similarity to the 2015 TX24. In the $216-219^\circ$ range one can find
many objects which similarities to the 2015 TX24  within $D'<0.015$,
reaching $D'=0.0055$ for the fireball of October 31, 2015 seen at 23:12
UT. A similarity to the 2005 TF50 is observed in the
$215-220^\circ$ range without a noticeable maximum. For the majority of
the Taurids from that range a simultaneous similarity to the 2015 TX24,
the 2005 UR and the 2005 TF50 seems to be typical; those objects move on
quite similar orbits but the 2005 TF50, as an object with a noticeably
higher orbit inclination, is distinctly less like the observed Taurids. 

If you compare Taurids to the 2003 UV11 and the 2007 RU17 at the
$\lambda_\odot=220^\circ$, the high similarity is noticeable but the
majority of objects similar to 2003 UV11 show also a very close
similarity to the 2007 RU17. The events observed during further nights
were characterized by similarity to the 2010 TU149 and the 2007 UL12.
Most of the phenomena observed around $\lambda_\odot=226^\circ$  have
orbits similar to the ones of those two objects.

For the 2005 TB15 there were only 15 objects with $D'<0.06$. The period
in which observed objects with similar orbits are situated is very wide,
encompassing practically the whole range of solar longitudes. For
$\lambda_\odot=219^\circ$ there were two meteors for which $D'$ compared
to the 2005 TB15 was 0.0236 and 0.0246, respectively. Taurids similar to
the 2005 TB15 differ distinctly from others when it comes to
different parameters. The average geocentric velocity for that group is
25.7 km/s that is about 5 km/s lower that in case of rest of the Taurids.
The mean geocentric radiant for this group is
$\alpha=49^\circ$ and $\delta=12^\circ$, whereas an average radiant for
all registered Taurids is $\alpha=51^\circ$ and $\delta=14^\circ$. 
Fig. 9 shows positions of radiants for objects similar to 2005 TB15. These are
meteors with the highest perihelion distance among all examined Taurids.

\begin{figure}
\centering
\includegraphics{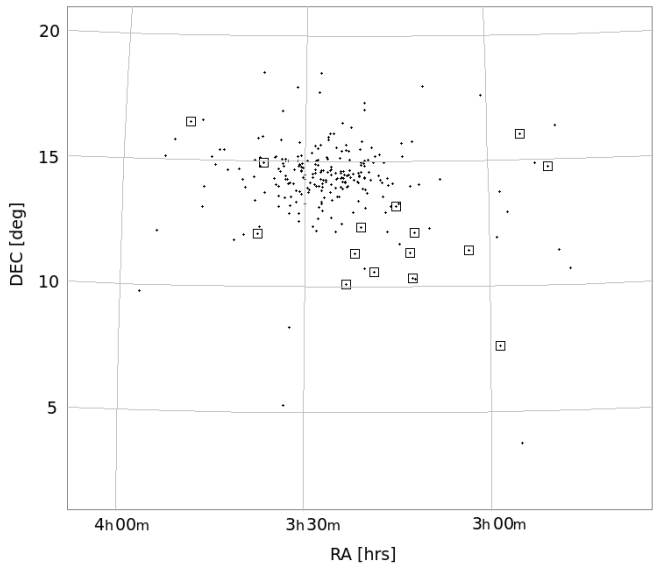}
\vspace{7.8cm}
\caption{Radiants of Southern Taurids, including radiants of the Taurids
showing similarity to the 2005 TB15 object (marked with squares).}
\end{figure}

\subsection{Comparison of activity in 2005 and 2015}

A big part of fireballs in 2005 and 2015 showed a very pronounced
similarity to the orbit of the 2015 TX24. That similarity is
especially clear for $\lambda_\odot=218^\circ$. The similarity of data
gathered in 2005 and in 2015 was checked separately. Despite the fact
that in 2005 we had definitely less data you still can perceive a
distinct peak, reflecting the similarity with the 2015 TX24. The
height and solar longitude of the peak for 2005 and 2015 fit each other
almost completely. It means the presence of the Taurids on orbits very
similar to the orbit of an object which orbital period is very close to
the 7:2 resonance with Jupiter. Fig. 10 presents that similarity very
clearly.

\begin{figure}
\centering
\includegraphics{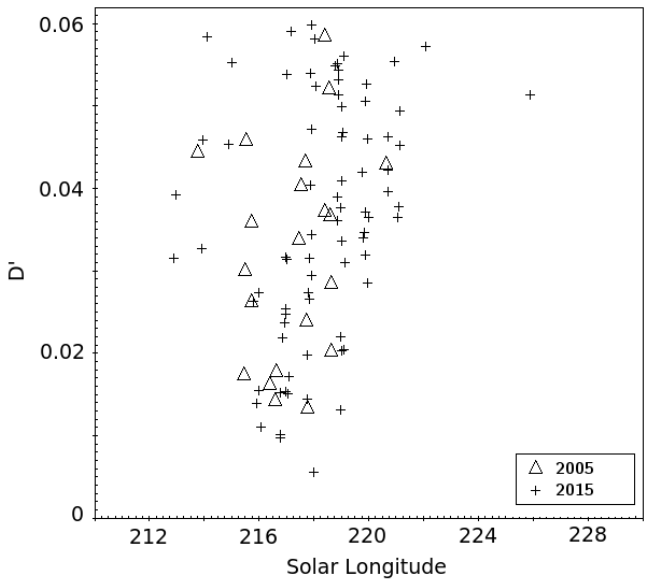}
\vspace{7.8cm}
\caption{Comparison of similarity of the orbital elements of Taurids to the 2015 TX24 object 
for 2005 and 2015 as a function of solar longitude.}
\end{figure}

Similarly for the objects connected to the 2003 UV11 and the 2007 RU17
a comparable activity was found in 2005 and in 2015. When it
comes to nights after November 4, in 2005 we had very bad weather
conditions and because of that it would be difficult to compare them
directly to respective nights in 2015; still for
$\lambda_\odot=229^\circ$ an activity from orbits similar to the 2010
TU149 was detected. Overall the activity pattern of 2005 and 2015 seems
to be very much alike. 

\section{Conclusions}

In this paper you can find a description of the activity of the
Taurids maxima observed in 2005 and 2015. Very bright events occurred
during both peaks of activity; especially in 2015 you could observe the
Taurids which brightness exceeded the brightness of Full Moon; there
were also many bright fireballs visible during several nights after the
maximum. For all Taurids observed in a period of 20 days starting at the
end of October and ending at the beginning of November 2005 and 2015 we
determined trajectories and orbits. The majority of Taurids' orbits is
similar to NEO objects which orbital periods which are in 7:2 resonance
with Jupiter. It concerns especially the 2015 TX24, the 2005 UR and the 
2005 TF50 as well as the 2010 TU149 which orbit is very similar
to orbits of the Southern Taurids observed at solar longitude of
$\lambda_\odot=226^\circ$. The NEO objects with orbits close to
resonance orbits have also perihelions distinctly different than the
perihelion of the 2P/Encke comet (noticeably smaller for the 2015  TX24,
2005 UR and 2005 TF50 objects, noticeably bigger for the 2010 TU149 and
the 2007  UL12 objects). These NEO objects should not be treated  as
parent bodies of the 7:2 filament, it is more correct to treat them as
the most massive remnants of the larger body fragmented  in the past. It
is worth to note that larger numbers of such bodies may exist in the
Southern Taurid stream creating a  kind of asteroidal core of the 7:2
stream.       

Among the observed Taurids the spread of orbital elements is quite even
no matter whether there are NEO objects in the stream or not. There is
very little chance of an accidental coincidence of those orbits with the
swarm of the Taurids and their presence is undoubtedly a result of a
long and complex evolution of the Taurid complex. Currently these are the
biggest objects circulating in the Taurid stream; what's more, the
list of NEO objects with orbits similar to those of the Taurids most
likely remains incomplete and it depends on the possibility of
detection from Earth.

This paper is another proof that the Taurid complex is
certainly one of the most interesting objects in the Solar System. It is
able to produce both impressive meteor maxima and extremely bright
fireballs (Dubietis \& Arlt 2006, Spurn\'y 1994) attracting the
attention of the media and ordinary people. Additionally, it can be
connected with catastrophic events like Tunguska (Kresak 1978, Hartung
1993) and can affect the climate on Earth (Asher \& Clube 1997).
Accurate observations and analysis of all kind of bodies associated with
the Taurid complex are then very a important task, demanding to continue
and affecting the safety of our planet.

\section*{Acknowledgments}

This work was supported by the NCN grant number 2013/09/B/ST9/02168.

\end{document}